# TinyLev Acoustically Levitated Water: Direct Observation of Collective, Inter-Droplet Effects through Morphological and Thermal Analysis of Multiple Droplets

*Adam McElligott, André Guerra, Michael J. Wood, Alejandro D. Rey, Anne-Marie Kietzig,*

*Phillip Servio\**

Department of Chemical Engineering, McGill University, Montreal, Quebec H3A 0C5, Canada

\*phillip.servio@mcgill.ca

ABSTRACT

Hypothesis: Understanding the crystallization of atmospheric water can require levitation techniques to avoid the influence of container walls. Recently, an acoustic levitation device called the TinyLev was designed, which can levitate multiple droplets at room temperature. Proximal crystallization may affect droplet phase change and morphological characteristics.

Methodology: In this study, acoustically levitated pure water droplets were frozen individually and in pairs or triplets using a TinyLev device. Nucleation, bulk crystal growth, and melting were observed using digital and infrared cameras concurrently.

Findings: Initially, the acoustic field forced the droplets into an oblate spheroid shape, though the counteracting force of the cooling stream caused them to circularize. Droplet geometry was thus the net result of streaming forces and surface tension at the acoustic boundary layer/air-liquid interface. Nucleation was determined to be neither homogeneous nor heterogeneous but secondary, and thus dependent on the cooling rate and not on the degree of supercooling. It was likely initiated by aerosolized ice particles from the air or from droplets that had already nucleated and broken up. The latter secondary ice production process resulted in multi-drop systems with statistically identical nucleation times. Notably, this meant that the presence of interfacial rupture at an adjacent droplet could influence the crystallization behaviour of another. After the formation of an initial ice shell around the individual droplets, dendritic protrusions grew from the droplet surface, likely seeded by the same ice particles that caused nucleation, but at a quasi-liquid layer. When freezing was complete, it was determined that the frozen core had undergone a volumetric expansion of 30.75%, compared to 9% for pure, sessile water expansion. This significantly greater expansion may have resulted from entrained air bubbles at the inner solid-liquid interface and oscillations at the moving phase boundary caused by changes in local acoustic forces. Soon after melting began, acoustic streaming, the buoyancy of the remaining ice, and convective currents caused by an inner thermal gradient and thermocapillary effects along the air-liquid interface, all contributed to the droplet spinning about the horizontal axis.

1. INTRODUCTION

Studying homogeneous nucleation is critical to understanding atmospheric chemistry and crystallization phenomena. This includes the formation of tropospheric ice particles as precursors to rain or as part of the cloud microstructure, as both can affect the wear on exposed parts of aircraft.[1-5] Unlike heterogeneous nucleation, where initial dendritic growth will occur at the solid-liquid interface from which freezing then proceeds[6-10], the homogeneous nucleation of atmospheric water begins by creating a shell of ice dendrites, usually over the entire air-liquid interface, before freezing towards the droplet center.[1, 11-16] **Figure 1** shows the different stages of droplet cooling in the context of changes in surface temperature over time. The first stage is supercooling, where the temperature drops from some initial value to below the equilibrium melting temperature ($T_m$). This stage continues until the nucleation time ($t_n$), at which point a critical radius is achieved. Note that it is possible for the droplet supercooling to reach the value of the environmental temperature ($T_{env}$), though this is not a requirement for nucleation.[1, 11, 14] Indeed, the nucleation of a solid phase is considered a stochastic process in supercooled liquid droplets: higher cooling rates can result in a significant range of nucleation temperatures.[4, 17] The second stage is recalescence, where the temperature rises to approximately $T_m$ due to the latent heat evolved in the liquid from the exothermic nucleation event and kinetic crystal growth of the shell around the droplet. This is a rapid stage, which results in significant internal pressure increases. These rises in pressure continue as the ice shell grows in the next stage and are often large enough to produce secondary ice particles (SIPs).[1, 11-13] In the third stage, the freezing stage, the temperature remains constant as there is a thermodynamic equilibrium between the heat released to the environment, via conduction, convection, and evaporation at the surface, and the heat produced through continued crystal growth. The fourth stage is ice-cooling, which begins when the droplet has fully solidified. It then cools to the environmental temperature. When this temperature is reached, there is no further heating or cooling and the droplet temperature stabilizes.[1, 11, 13, 14]

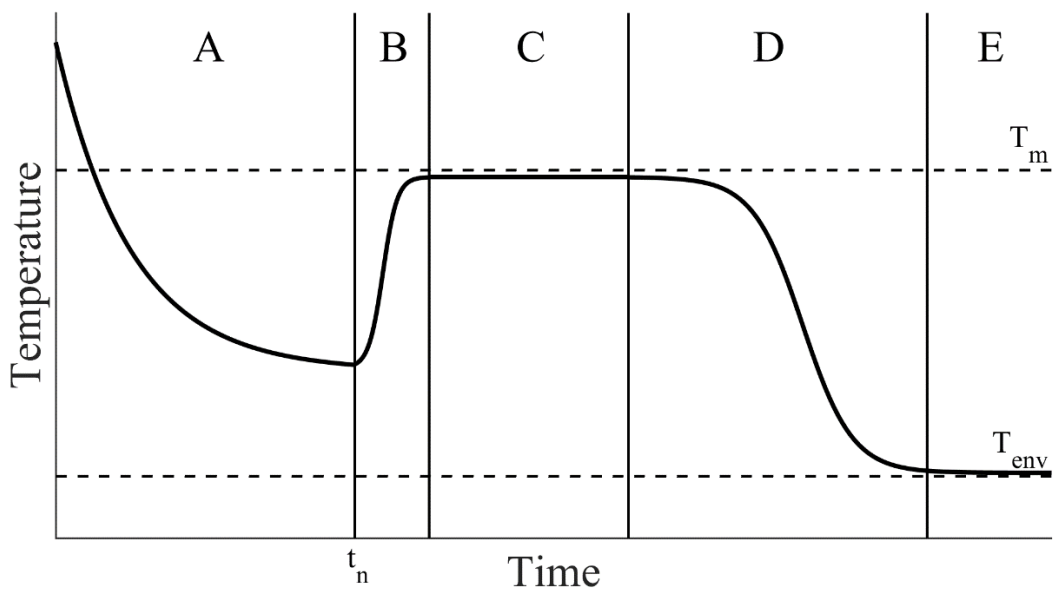

**Figure 1.** Temperature stages during the freezing of supercooled droplets. The first of these is supercooling (A), a metastable stage. This stage is followed by recalescence (B), freezing (C), ice-cooling (D), and stabilization (E). $T_m$ and $T_{env}$ are the melting and environmental temperatures, respectively. The time of the initial nucleation event is marked as $t_n$.

To study both homogeneous and secondary nucleation events, which is to say initial nucleation events caused by aerosolized ice particles already present in atmospheric air, several droplet levitation methods have been developed to eliminate the effects caused by the presence of solid surfaces and container walls while keeping a droplet in a stable position.[3, 18] Notably, there are non-obtrusive methods where the droplet surface is solely in physical contact with a surrounding cooling fluid, commonly air, while held against gravity. These can include optical[2], electrodynamic[19, 20], or free-fall[12] levitation. The latter method uses a laminar flow of moist, cold air to hold the droplet in place, which mimics free-fall at terminal velocity.[12] There is also acoustic levitation, where an acoustic transducer, for example, a Langevin horn, produces sound waves that bounce off a concave reflector. This configuration generates a standing wave that traps small particles at its nodes on a single axis.[21] Using these types of levitation, single-droplet studies have been conducted for water, including examinations of nucleation and freezing[3, 14, 16, 20], structure and surface tension[22, 23], evaporation[24, 25], and gas-liquid interfacial flow[26]. Moreover, these studies have moved beyond the physicochemical properties and crystallization dynamics of water and examine immersion freezing techniques[5], multi-component analyses[27-29], gas hydrate formation[30, 31], and new tools and analytical methods such as trace analysis and breakdown spectroscopy.[32-34] Building upon this extensive base of research, a logical next step is to examine multiple levitated water droplets simultaneously, as lone droplets rarely exist in nature.

Recently, an acoustic levitation device called the TinyLev was developed. It is a single-axis, non-resonant levitator that uses multiple ultrasonic transducers to produce a stable trapping force that is robust to temperature and humidity changes, though freezing temperatures were not examined.[21] The TinyLev has several advantages in droplet freezing studies over other acoustic levitation methods. Chiefly among these is its ability to levitate several droplets on a single axis simultaneously. Moreover, the device has an open configuration and is not contained within a cooling chamber. This means that one may control the droplet size and choose the node on which the droplet is placed. Furthermore, this configuration allows for direct measurements: digital and infrared cameras can be pointed directly at the droplets for data acquisition with no obstructing intermediate surface. More information on the TinyLev can be found in Marzo et al. (2017).[21]

This study will examine the morphological and thermal behaviour of water droplets suspended in an acoustic field during phase change from liquid to solid ice, as well as melting. Multiple droplets will be frozen simultaneously, up to three, to determine the presence of inter-droplet effects, if any. Additionally, the phase change will be under direct observation of two synchronous cameras. This study will focus on the complex interaction between geometry, phase transition, and transport processes. It will examine the interplay between conformational changes in droplet shape, capillary processes, as well as heat, mass, and momentum transport, which are all present during crystallization and melting. The factors affecting droplet nucleation, and the dominant nucleation mechanisms, will be duly explored. To the best of our knowledge, this is the first time that the effects of multiple, concurrent water droplet crystallizations have been examined in a levitation device. Furthermore, this study will also focus on droplet morphology beyond the nucleation and initial freezing stages, which have been the focus of most previous studies, and will examine protrusion formation, bulk crystal growth, and melting. Unobstructed, direct measurement of this type of system with coincident cameras is also entirely novel.

2. MATERIALS AND METHODS

A simplified schematic of the experimental setup described in this section is provided in **Figure 2**. The TinyLev consists of two arrays of 36 Murata 10 mm acoustic transducers operating at 40 kHz and arranged in rings. Instead of a transducer, the center of each array is a hole vertically aligned with the axis of levitation. Furthermore, the frame housing the arrays is 3D printed as a single piece, and the upper and lower array surfaces are spaced 7.5 cm apart. This results in trapping forces of approximately 0.1 laterally and 0.5 longitudinally, relative to the input power. This configuration creates eleven nodes in which droplets may be placed.[21] In the case of the crystallization experiments conducted for this study, it was possible to freeze five droplets simultaneously. However, capturing the phase change of this many droplets did not give sufficient resolution for analysis. Therefore, only the three center droplets, which are highlighted in **Figure**

**2**, were examined in this study. Note that these are numbered in terms of position, where the top node is designated as position 1, followed by position 2 in the center, and position 3 at the bottom. The average room temperature was 21.50 ± 0.55 °C, and the average humidity was approximately 32.00 ± 5.60 %. The TinyLev was powered through a Delta Elektronika power supply (SM 70-AR-24) operating at 10 V and 0.8 A. Usually, 11 V is required for water levitation, which is a value specific to its density.[21] However, the density of ice is less than that of water, so only 10 V was required to ensure droplet stability throughout freezing and melting. The power runs through a driving board consisting of an Arduino Nano, which generates square wave excitation signals, and an L297N Dual H-Bridge motor driver, which amplifies the signals. Further details on the TinyLev and accompanying driving electronics are available in Marzo et al. (2017).[21]

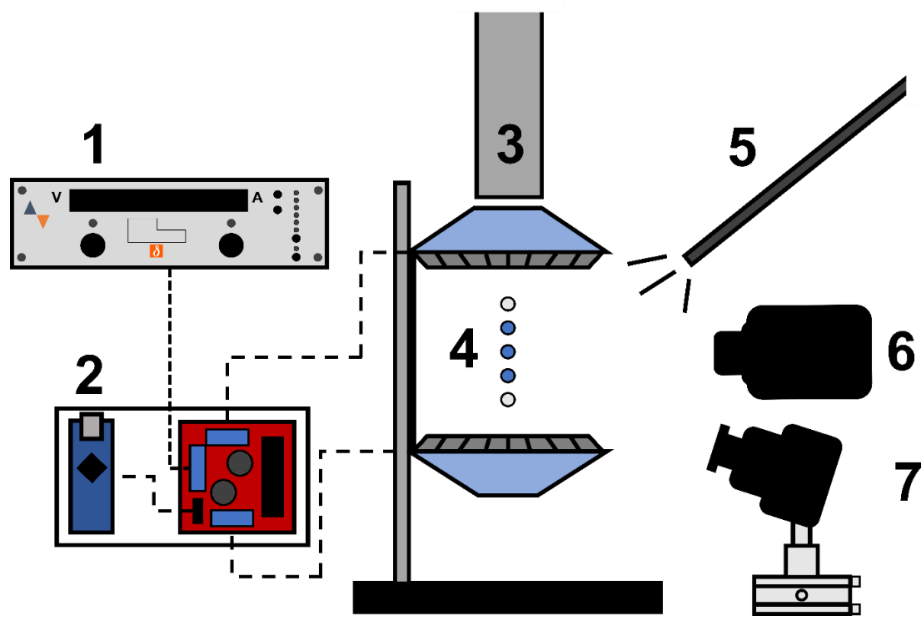

**Figure 2.** Simplified schematic of the experimental setup. This includes a power supply (1), driving board with Arduino and amplifier (2), cryogun (3), TinyLev acoustic levitator with black backing (4), LED light source (5), infrared camera (6), and camera with lens and multi-axis manual positioner (7). Note that during use, the lines of sight of the infrared and digital cameras are perpendicular, with the digital camera pointed directly at the levitator backing and parallel with the table. Their orientation in this figure is simply for clarity.

The droplets were frozen using a custom-built stainless steel 316 cryogun. The cryogun is a 2.5 cm diameter cylinder, and its walls and bottom are 2 mm thick. This cylinder was placed directly above the opening of the top array such that the distance from the end of the cryogun to the levitated droplets was 6.4, 6.9, and 7.4 cm for the top, middle, and bottom positions, respectively. The freezing works by filling the cylinder with 35 mL of liquid nitrogen. The temperature at the surface of the bottom of the cylinder stabilizes to approximately -55.5 °C. This cools the air directly below the cylinder, which increases in density and falls due to gravity. This creates a significant cooling stream around the droplets maintained through each run. Levitational systems often have weaker airflows compared to free-fall levitation techniques so as not to destabilize the droplet.[5] The airflow created in the system used for this study is at the velocity limit: occasionally droplets were destabilized by the strength of this airflow and the experiment was restarted. Images of the droplets were taken with a Canon EOS 60D DSLR camera (18.0-megapixel CMOS sensor) equipped with an MP-E 65 mm f/2.8 1–5x macro lens. The camera was mounted onto an OptoSigma multi-axis manual translation stage to fine-tune the optical focal point. The camera was placed normal to the axis of levitation, so parallel to the table, and the lens was 4.5 inches (11.43 cm) away from the axis for single-droplet experiments and 2.5 inches (6.35 cm) for multi-droplet experiments. Droplet illumination was from a fibre optic LED and a black backing was placed in the TinyLev for more precise visualization. Thermal images were captured using a Jenoptik IR-TCM 384 infrared camera calibrated for a 1-2% measuring accuracy in the range of -20 to 20 °C and with a NETD temperature resolution of less than 0.08 °C. This camera has its lens placed 6 inches (15.24 cm) from the axis of levitation and its line-of-sight perpendicular to that of the DSLR. The VarioCapture 2.7 software was used to obtain infrared images.

  Before data acquisition, power was supplied to the TinyLev for droplet placement. A BD 3 mL syringe was used to place single droplets in any of the three positions or four positional configurations: positions 1 and 2, 1 and 3, 2 and 3, or 1, 2, and 3. The water had previously undergone reverse osmosis (RO) treatment with a 0.22 μm filter. It had a total organic content of less than 10 ppb and a conductivity of 10 μS. Furthermore, the cryogun was filled with liquid nitrogen, and a minute was allowed to elapse to ensure the surface of the cryogun had sufficiently cooled and a steady air stream was produced. Note that a layer of frost would often form on the sides of the cryogun, and thus after the elapsed time, the edges would be scraped of frost up to 2 cm from the cylinder bottom. Additionally, after the one-minute delay, data capture from both cameras was initiated, and the cryogun was placed above the TinyLev opening, which created a cold air stream over the present liquid sample(s). To ensure complete freezing of the droplet and to examine post-solidification events, the cooling period was set to three minutes, after which the cold air stream was removed, and the droplets allowed to melt. The melting period was set to one minute. Fifteen replicates were performed for each position or position combination. This was considered sufficient due to the high degree of undercooling, which led to a high likelihood of nucleation and similar conditions between runs. Note that an additional 30 replicates were performed for the single droplet at position 2 (for a total of 45) such that an accurate droplet size distribution could be obtained and generalized to each position. Also, note that melting experiments were only performed for these 45 replicates as, unlike the cooling rate, the heating rate is the same for all droplets, and it was not believed that multiple droplets would affect each other during melting. Therefore, 135 replicates were performed, of which 45 included melting, and analyzed for this study.

3. RESULTS AND DISCUSSION

Water droplets were suspended in an acoustic field and frozen using a cryogun. These were either singular droplets, two droplets occupying any combination of the three available positions, or three droplets occupying all available positions. **Figure 3** shows the morphological and thermal behaviour during the crystallization process. The characteristic behaviour shown in this figure was exhibited by all droplets in some form, regardless of position or droplet number, and is divided into six parts. Part A is the droplet at rest prior to the placement of the cryogun. Part B is circularization, where the cryogun is placed, cooling begins, and the droplet shape moves from an oblate spheroid to a more spherical shape (i.e., a shape with lower eccentricity). In this and the remaining thermal images, the background is a darker shade of pink, which shows the presence of a cold stream. Part C is nucleation. Notice that there is an increase in temperature due to the exothermic nucleation event. This corresponds well with the stages of droplet crystallization described in the introduction. Part D is the initial formation of protrusions at the droplet's surface, while Part E is the bulk growth of ice within the droplet and the elongation of the protrusions. The final part (F) is complete crystallization of the droplet when there is no liquid water present, and the droplet reaches its lowest measured temperature. The effects demonstrated in **Figure 3** are expanded upon and explained in the following sections, note that in (F) the droplet appears smoother in the IR image due to its fast rotation.

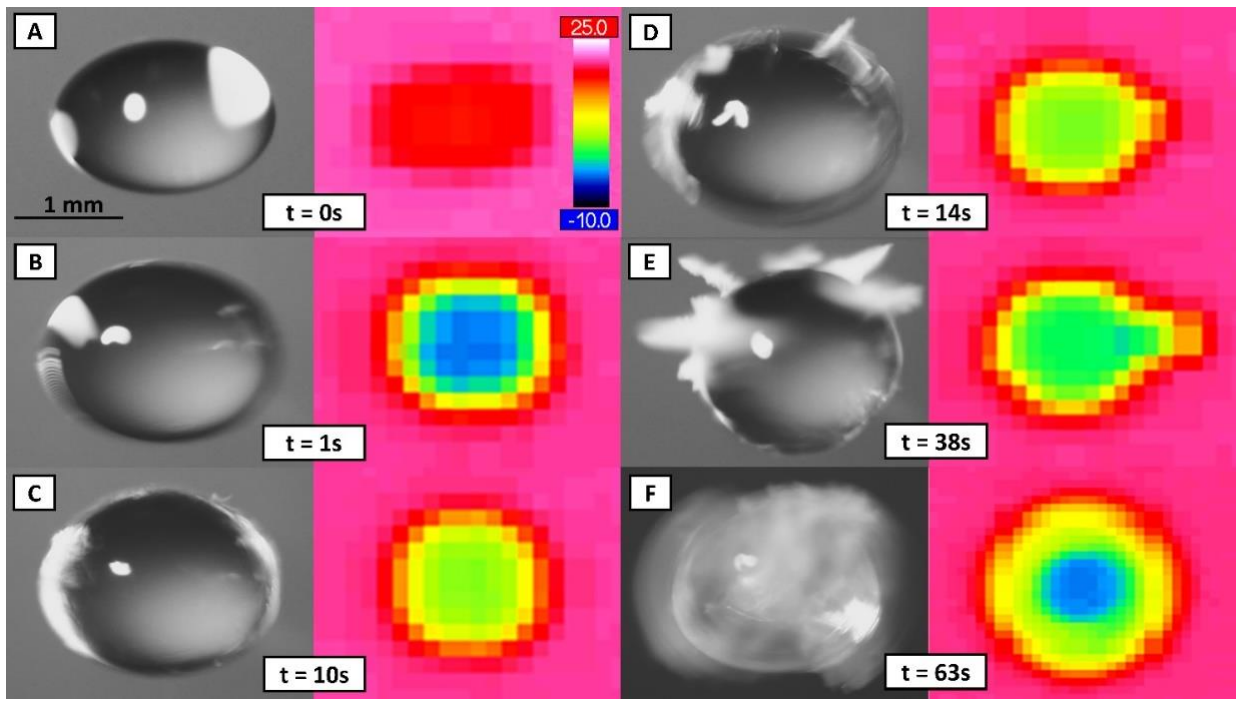

**Figure 3.** Morphological and thermal images of crystallization in a single drop. These include the droplet at rest (A), circularization (B), nucleation (C), initial protrusion formation (D), bulk growth and protrusion elongation (E), and complete freezing (F). Note that due to the resolution of the IR camera, the thermal images are not to scale.

3.1 Suspended Water Droplet to Nucleation

3.1.1 Initial Conditions, Droplet Stability, and Circularization

Before the cryogun was placed and cooling began, droplets were levitated in a combination of three positions in the acoustic field. In **Figure 3**A, it is evident that the droplets do not adopt a perfectly spherical shape but instead that of an oblate spheroid, where the horizontal radius is greater than the vertical one. This shape is likely caused by the forces present within the acoustic field. Sound is a mechanical wave, and so it carries momentum, which generates acoustic radiation forces. Moreover, high levels of acoustic pressure are required to counteract the gravitational field and keep the droplet stabilized in the air.[35] As significant pressure gradients are produced over the droplet's surface, this results in an appreciable net force. These body forces prevent the droplets from migrating out of the acoustic region and distort the drops.[22, 23, 26, 36] These forces are visualized in **Figure 4**A. The figure shows inner and outer acoustic streamlines at the acoustic boundary layer just before the droplet surface caused by high amplitude oscillations.[25] The acoustic streamlines create circular vortices at the upper and lower parts of the droplet and determine the droplet boundary.[28]

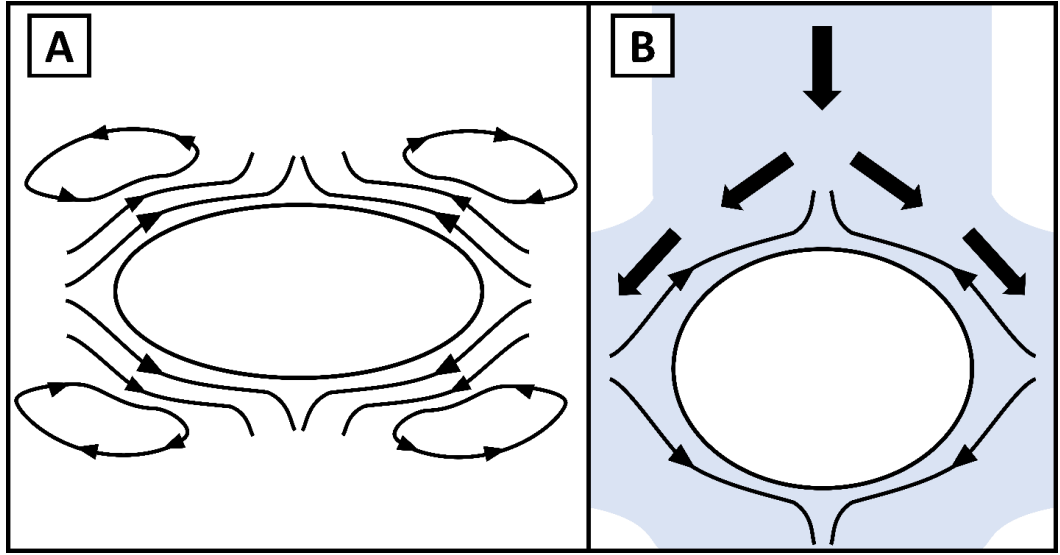

**Figure 4.** The shape of levitated water droplets varied between an oblate spheroid (A) formed from acoustic streaming forces and a close to spherical shape (B) when the cooling stream reduces the field strength at the acoustic boundary layer.

The oblate spheroidal shape, however, is characteristic of natural atmospheric droplet shapes, specifically of freely falling raindrops.[5, 15, 37] Moreover, as the droplet is surrounded solely by air, the conduction of the latent heat released from the drop during freezing is also similar to atmospheric conditions.[5] The average initial temperatures of the droplets, around 17.5 °C, are lower than the room temperature of 21.5 °C and decrease moving downward: from 17.94 ± 0.49 °C at position 1 to 16.83 ± 0.61 °C at position 3. Previous studies have found temperature deviations of about 1 °C at the nodal planes caused by acoustic streaming.[37] Acoustic streamlines cause the water to flow steadily at the surface, resulting in convective cooling through small amounts of evaporation. In turn, the temperature at the surface is reduced, which may explain why levitated droplets have initial temperatures that are cooler than the surroundings.[36]

Note that the above-described effects are present in all droplets regardless of position. However, there were some positional differences. It has previously been found that the most stable node is spaced equally between two sources of acoustic waves as the longitudinal forces are more symmetrical there compared with the other positions.[22, 37] This may mean that, while the middle three drops are trapped with the most similar amplitudes, droplets placed in position 2 may have oscillated less than in positions 1 or 3, which could have had more significant radial oscillations.[21] Furthermore, droplets in positions 1 and 2 have similar average volumes ($1.35 \pm 0.31$ µL), whereas those in position 3 have much greater volumes ($2.10 \pm 0.47$ µL). Position 3 droplets may be larger because there is a greater acoustic force on them due to their proximity to the stronger lower array, which allows them to be larger without fragmenting. The lower array necessarily has a stronger acoustic force as it creates pressures that counteract the force of gravity. As larger droplets have larger surface areas, this would also result in greater acoustic streaming at the droplet and slightly more evaporation at the surface. While this may not have affected the equilibrium volume of the droplets, it could explain why the average initial temperature decreases as the positions lower.

When cooling was initiated and the stream surrounded one or more droplets, all individual droplets had significant reductions in eccentricity. This is visualized in **Figure 4**B. The cold gas flow provides a uniform lateral force on the sample, which may provide sufficient pressure to reduce the effects of acoustic streaming.[22] The shape of the droplet is a function of the balance between the normal forces acting upon it, which were the acoustic field and gravity, and the surface tension and hydrostatic forces within the droplet. The addition of a downward force likely modified this net force and reduced the droplet's eccentricity.[2, 34] This shape is not a complete sphere; the droplet remains an oblate spheroid (or ellipsoid), so it is still akin to freely falling raindrops, simply falling at a higher velocity. It is possible that the internal viscous flow patterns also change from oblate to more circular in this period, though this was not detected in this work and may be subject to future investigation.

### 3.1.2 Nucleation: Time, Temperature, and Cooling Rate

The onset of nucleation was marked by two simultaneous phenomena, as shown in **Figure 3**C. The first was the formation of a dense, dendritic ice shell around most or all of the air-liquid interface. The second was an exothermic release of heat where the temperature rose to the equilibrium melting temperature. Both of these phenomena have been observed previously.[1, 11-14, 16] The formation of the primary ice shell is considered adiabatic, as it is quick enough that heat is solely stored in the droplet and not released to the surrounding air.[16] In levitated crystallization, the droplets are often stagnant in the acoustic field, and thus the transport of latent heat to the environment is not uniformly distributed at the droplet surface. Instead, there is a local maximum of heat exchange at the top of the droplet and a minimum at the bottom. This can create weak spots or bulging of the ice shell.[11-14] However, the presence of an airflow causes the droplet to rotate, usually on an axis normal to the flow direction, and results in more uniform cooling.[12] The time required for nucleation, the cooling rate at each droplet, and the droplet temperature at the nucleation time are provided in **Table 1**. The cooling rate is simply the change from the initial temperature to the nucleation temperature divided by the nucleation time. At longer time scales, this value can be reduced as temperature decreases (see **Figure 1**). However, in the time frame considered in this study, it can instead be considered constant and was measured to be so. The position 1 droplets, which were closest to the cryogun, had the fastest nucleation times and the fastest effective cooling rates. This meant that the position 1 droplets had faster primary shell growth than those in the other two positions and, as a result, more opaque ice shells.

**Table 1.** Nucleation time, effective cooling rate, and nucleation temperature for droplets at each position in each configuration. Note that the 95% confidence intervals on the nucleation time and nucleation temperature are provided in brackets (±).

|  | Individual | Positions 1/2 | Positions 1/3 | Positions 2/3 | Positions 1/2/3 |
|---|---|---|---|---|---|
| **Nucleation Time (s)** | | | | | |
| **Position 1** | 8.76 (0.69) | 8.93 (1.69) | 9 (1.14) | - | 9.33 (1.10) |
| **Position 2** | 12.6 (2.56) | 11.53 (2.70) | - | 10.53 (1.55) | 10.6 (1.76) |
| **Position 3** | 11.07 (1.64) | - | 11.53 (1.23) | 11.2 (1.02) | 11.33 (1.75) |
| **Cooling Rate (°C/s)** | | | | | |
| **Position 1** | 2.23 | 2.18 | 2.21 | - | 2.11 |
| **Position 2** | 1.45 | 1.72 | - | 1.84 | 1.83 |
| **Position 3** | 1.68 | - | 1.74 | 1.72 | 1.71 |
| **Nucleation Temperature (°C)** | | | | | |
| **Position 1** | -1.59 (0.31) | -2.02 (0.45) | -1.52 (0.33) | - | -2.00 (0.20) |
| **Position 2** | -1.08 (0.47) | -1.73 (0.29) | - | -1.85 (0.14) | -1.75 (0.07) |
| **Position 3** | -1.80 (0.10) | - | -1.75 (0.07) | -1.75 (0.07) | -1.75 (0.06) |

Homogeneous nucleation rates are not dependent on droplet volume, but solely on the level of undercooling, which increases the level of supersaturation.[16, 38] For instance, we see that individual droplets in position 1 both reach colder temperatures and nucleate faster than individual droplets at position 2. However, individual droplets at position 3 reached the coldest temperatures before nucleating compared to the other positions but nucleated more slowly than droplets at position 1. Moreover, the nucleation of ice in levitated droplets is surface-dominated, and acoustic levitation conditions already enlarge the droplet surface area.[36] In turn, this can increase the droplet's propensity to be supercooled. Therefore, particles with greater surface areas have the potential for more active sites for ice nucleation, which means the probability of freezing is proportional to particle surface area.[5] In fact, dendritic freezing of supercooled droplets is estimated to have a greater than 90% probability of occurring on the surface.[16] Therefore, the coldest droplet with the highest surface area (position 3) should see the fastest nucleation times. However, the data suggest that the nucleation times depend more on the cooling rate than on their undercooling, which indicates that nucleation in these droplets is likely not homogeneous.[5]

Recent studies have provided strong evidence for heterogeneous nucleation. Tabazadeh et al. (2002) suggested that the most thermodynamically favoured mechanism for nucleation in atmospheric droplets is pseudo-heterogeneous and occurs at the air-liquid interface (the droplet surface).[39, 40] From this point, the ice would propagate from the surface into the bulk.[39] This behaviour is observed in droplets under cooling streams[13] and was also observed in this study. Furthermore, due to droplet circularization, the shape of the liquid phase prior to nucleation is close to the equilibrium shape for this kind of nucleation (from the Wulff construction).[41] Additionally, ultrasonic waves have previously been found to promote heterogeneous ice nucleation.[18, 36, 42, 43] Ice formation in levitated droplets is characterized by more numerous nucleation sites with higher nucleation work compared with systems using solid-containing surfaces.[31] However, entrained microbubbles near the air-liquid interface can act as heterogeneous nucleation sites, reducing the nucleation work at the most undercooled regions of the droplet.[30] Acoustic pressure concentrates near the droplet's surface, which can cause a cavitation effect and increase the number of microbubble nuclei.[18, 36, 42, 43] In turn, droplets in an acoustic field become more likely to nucleate heterogeneously. All these effects may be strongest at position 3, which is not only near the lower array but has the highest surface area. However, position 3 again does not have the fastest nucleation time, so primary nucleation mechanisms may not be the cause of nucleation.

Instead, it is highly possible that secondary nucleation caused by subvisible aerosolized ice nucleation particles (INPs) is the dominant nucleation mode. It is estimated that small ice aerosol particles cause more than 90% of nucleation events in levitated droplets.[16] Primary nucleation mechanisms could still be present for some droplets, but they occur at significantly lower rates. Aerosol particles from water in the air can be supercooled easily due to their small volume.[4] The system has a cryo-stream but is also subjected to room temperature, creating a significant temperature gradient in the gaseous environment. When these gradients are present in a stationary ultrasonic field, it has been shown that water or ice aerosol gathers at the pressure nodes to displace the warmer gas mixture where the droplets are present.[15, 44] At an empty node, these can even form new visible ice or snow particles, which was observed in this system. There are three stages to the formation of particles at empty nodes: the agglomeration of individual aerosol particles due to the presence of a quasi-liquid layer on the ice particle surface, the accumulation of aerosol agglomerates, and further particle growth via gas-phase deposition.[15, 44] Unlike the rotational ice shells of the droplets, which may not occur in nature, the new particles demonstrate shapes that are found in nature which suggests that the influence of the acoustic field on their formation is small; the field only concentrates their location.[2, 15, 44] Therefore, in the same acoustic field, secondary nucleation at any position should be a function of gas stream temperature, and so the distance from the cryogun and cooling rate. However, the single position 3 droplet furthest from the cryogun has the second-highest cooling rate. It may be that the stronger acoustic field at that position is causing more aerosolized ice to be directed there. Additionally, the position 3 droplets have a greater surface area on which secondary ice particles could collide. One or both effects would reduce the nucleation time and effectively increase the cooling rate, so the secondary nucleation time is not a function of undercooling. Therefore, as nucleation time is a function of cooling rate, which strongly promotes nucleation via aerosolized ice particles, the data suggest that the nucleation in these droplets is mainly secondary and not primary.

The above-described effects are always present regardless of droplet configuration. If they were the only present effects and there were no effects from having multiple droplets present, then the nucleation times would not change in the multi-droplet configurations. However, when there are droplets in positions 1 and 2, or positions 2 and 3, the nucleation time at position 2 decreases (though within error) from the single-droplet value such that there is no statistical difference between the nucleation times at the two positions. In other words, when droplets are in proximity to each other, statistically, they nucleate together (i.e., individual droplets have statistically different average nucleation times, systems with adjacent droplets do not). It should be noted that, when examining **Table 1**, it appears that in the 2/3 system position 2 has a lower nucleation time than position 3, but only position 2 has a change in nucleation time. However, because the averages are not statistically different, it cannot be stated that position 2 necessarily nucleates first, only that the droplets nucleate together. In fact, as the time at position 3 remains constant, nucleation at that position may be affecting position 2 even if the average is lower. When droplets are in positions 1 and 3, there is no significant change in the nucleation times at either position: the 1 cm distance between them may be large enough for them not to affect each other (i.e., INPs from position 1 may become trapped at position 2 and would not affect position 3). However, when there are droplets in all three positions, the nucleation time at position 2 decreases again, and, critically, there is no statistical difference between the nucleation times at any position. There may therefore be additional effects when multiple droplets are present. After the initial dendritic formation of a primary ice shell, the droplet's core remains primarily liquid.[11, 16] The specific volume of ice is greater than that of water, so high pressures are induced when the ice shell begins to grow, imposing significant mechanical stress. This stress can cause the ice shell to deform or crack, the latter of which results in the emission of secondary ice-nucleating particles (INPs).[3, 11] The likelihood of INP production in droplets with diameters larger than 300 μm, such as the ones in this study, exceeds unity.[11] These INPs may then be directed toward the next droplet position by the thin necks between droplets created by the acoustic field.[44] The secondary ice formation mechanisms present in the system are likely cracking and breakup. Cracking occurs when the pressure inside the droplet exceeds the mechanical stability of the surrounding ice shell. This causes large cracks to form, and small ice particles are emitted in the process. Breakup occurs when the cracking has sufficient energy to overcome capillary forces and negative pressure inside the droplet, and it fragments into two hemispheres perpendicular to the crystallographic axis.[1] Again, INPs are ejected when this occurs.[11, 38] The occurrence of cracking and breakup is due to the presence of the cooling stream. The presence of this stream induces droplet spinning and results in a more even ice shell density. Moreover, the more efficient latent heat transport caused by forced convection enhances the rate at which the ice shell grows. The mechanical stress on the shell also increases faster, and thus there is a high fragmentation frequency.[12] This observation is consistent with free-fall droplet freezing experiments, which found that nearly half of the droplets experienced breakup: a full order of magnitude more frequent than a droplet levitated in stagnant air.[1, 12, 13] Therefore, the decreases in the nucleation times at position 2 and lack of statistical difference between these times in multi-drop systems may be due to secondary nucleation particles produced during the post-nucleation period of already-nucleated droplets. When the nucleation of one droplet produces these particles, they are directed towards the droplet below, increasing the number of INPs at that position and thus the likelihood of nucleation.[13] Note that there may also be no effect in the position 1/3 system if these particles become trapped in the position 2 node between the droplets.

### 3.2 Protrusion Formation to Complete Freezing

#### 3.2.1 Protrusion Time and Morphology During Bulk Freezing

After nucleation, the ice shell began to grow inwardly. However, growth also occurred on the outer surface of most of the droplets and several protrusions formed. Images of the different kinds of protrusions observed throughout the study can be found in the supporting information. These included thin, wisp-like protrusions which could surround the droplet, thick protrusions which formed a compact band or ring around the crystallographic center, or multiple lower-density rings at different locations on the droplet. Protrusions were therefore often ring-like due to the spinning of the droplet, though individual spike-like protrusions could form as well. There was no evidence to suggest that droplet position affected protrusion formation. Moreover, protrusion formation did not affect the stability of the droplets; the same droplet could remain in position despite having flipped over. The shapes of these droplets may have been affected by random droplet oscillations caused by the acoustic forces present. Likewise, local changes in acoustic wavelength are created when sample temperatures begin to deviate substantially from surrounding gas temperatures.[22] Local distortions allow the particle to be rotated, usually around arbitrarily oriented axes.[44] It is also possible that colder droplets develop particular protrusions compared to warmer ones due to acoustic wavelength differences. However, droplet temperatures did not differ by more than 3.00 °C, even for different runs at the same position that could exhibit various protrusion types. This indicates that any temperature effects not significantly influence protrusion shape.

It is possible then that INPs, the same from the previous section, which came from the water in the air or particle nucleation, also cause the initial formation of a protrusion. Pressure changes and heat evolution could result in minor remelting of the outer part of the ice shell, which may create a quasi-liquid layer on the droplet exterior.[11, 15] Discrete aerosolized ice particles could impact this layer one or multiple times and become the root of a protrusion. Droplet oscillation and the further agglomeration of other particles, combined with the droplet's rotational motion, could produce the different protrusion shapes. The time from experiment initiation to the formation of the first protrusion is presented in **Table 2**. As all three positions are subject to the same particle concentration from the air, these times may depend on the increase in particle concentration from secondary ice production at nucleation. These particles are too small to be detected optically, which imposes a significant uncertainty on their size and number. However, it is possible to calculate the instantaneous change in pressure at nucleation, which is an indicator of INP production.[13] The change in the freezing point pressure is calculated through the Clapeyron equation:

$$\Delta p = \Delta T \left(\frac{\partial p}{\partial T}\right)_{s,l} \tag{1}$$

where $\Delta T$ is the freezing point depression at the phase boundary in Kelvin and $\left(\frac{\partial p}{\partial T}\right)_{s,l}$ is 134.6 bar K$^{-1}$ for water at 1 bar and 0 °C.[45] The freezing point depression requires the melting point and the temperature at the interface, but only the droplet surface temperature was measured. Therefore, a model constructed by Kleinheins et al. (2021) was used to find the temperature at the phase boundary using the measured cooling rate and a time correction factor:

$$T_{pb} = T_1 - b\delta^{-1} \quad (2)$$

$$\delta = \frac{b}{2(T_1 - T_2)} \left[ \sqrt{1 + 4\frac{T_1 - T_2}{kb}} - 1 \right] \quad (3)$$

where $T_1$ is the droplet temperature in Kelvin (K), b is the measured cooling rate in K/s, $\delta$ is a correction factor to determine the net temperature deviation for a droplet in continuous airflow[5] in $s^{-1}$, $T_2$ is the measured nucleation temperature in Kelvin, and k is a constant of 1.8 $s^{-1}$.[13] The results of these calculations are found in **Table 2**. Note that these internal pressure changes are a function of the measured cooling rate, which is only an effective rate. Therefore, they may be overestimations of the actual changes but are still in the correct order of magnitude and relevant for comparison.[13]

**Table 2.** Time required for the first protrusion to form, and internal pressure change immediately after nucleation for droplets at each position in each configuration. Note that the 95% confidence intervals on the protrusion time are provided in brackets (±) and none are provided for the pressure change as they are only approximations.

|  | **Individual** | **Positions 1/2** | **Positions 1/3** | **Positions 2/3** | **Positions 1/2/3** |
|---|---|---|---|---|---|
| Protrusion Time (s) | | | | | |
| **Position 1** | 22.12 (4.39) | 25.20 (7.98) | 31.53 (5.40) | - | 25.00 (3.36) |
| **Position 2** | 26.91 (4.95) | 28.27 (8.30) | - | 26.80 (5.78) | 30.07 (5.44) |
| **Position 3** | 25.73 (8.07) | - | 32.33 (7.00) | 23.73 (4.31) | 32.40 (9.31) |
| Internal Pressure Change (bar) | | | | | |
| **Position 1** | 490 | 455 | 488 | - | 436 |
| **Position 2** | 316 | 349 | - | 374 | 378 |
| **Position 3** | 336 | - | 355 | 349 | 347 |

An examination of the values in **Table 2** supports the hypothesis that protrusions are caused by secondary nucleation. However, due to significant variations in protrusion time, it should be noted any correlation is likely weak. However, looking at the individual droplet positions, the time to the first protrusions correlates to the internal pressure change at nucleation. Higher internal pressures may result in more production of INPs.[13] In turn, droplets with higher internal pressures have a greater local concentration of ice particles at their surface from the INP production mechanisms. These particles can be redirected towards the droplet by the acoustic field and form protrusions. In summary, the protrusion time is likely internal pressure-dependent, so it is also INP concentration-dependent, and therefore protrusion formation likely begins with a secondary nucleation event. Furthermore, droplets in position 1 always had the highest internal pressure changes and the shortest initial protrusion times in multi-drop systems. This correlation follows for the position 2 and position 3 droplets in these same systems, except in the position 2/3 system. This may be because this system has the smallest difference in pressure change between droplets and thus similar INP concentrations. Specifically, assuming isotropic elasticity, the strain on the ice shell likely depends only on the stiffness for a given pressure load and on the inner and outer radii; there may be no elastic contribution. Additionally, the Tresca yield criteria for shell buckling is:

$$|\Delta P| = \frac{2\sigma_0}{3}\left[1 - \left(\frac{r_o}{r_i}\right)^3\right] \qquad (4)$$

where $\sigma_0$ is the yield constant, and $r_o$ and $r_i$ are the inner and outer radii, respectively. When equation (4) is violated, a crack emerges. This equation defines the mechanical stability of the shell as a function of pressure load and shell thickness. Therefore, in the position 2/3 system, where both droplets have similar pressure changes, the ratio of the inner and outer radii may also be similar, even if the droplets in position 3 are larger. Hence, the 2/3 droplets have similar yield constants, cracking rates, and so similar local INP concentrations relative to their volumes.

In terms of size, the protrusion length from the ice shell could reach as large as 1.44 mm, and there was a minimum of 0.08 mm if they formed. On average, the protrusion length was approximately 0.66 ± 0.20 mm. However, in the triple-droplet system, the average was 0.91 mm. It is possible that the more droplets there are present in the system, the more INPs are also present; this could increase the number of protrusions and their size. However, long and short protrusions were observed at all positions. This may indicate that unpredictable oscillations largely control protrusion length. Specifically, oscillations in a particular direction could cause the accumulation of aerosolized ice or feed the protrusion using part of the quasi-liquid layer.

Additionally, different sorts of water beads could form. These beads formed for several droplets in addition to other protrusions and an image is provided in the supporting material. Again, there was no evidence to suggest that beading was position-dependent; they appeared at all positions with similar frequency. Furthermore, there was no greater or lesser occurrence of beading in multi-drop systems. Previous studies have indicated that "frozen noses" can form after nucleation from liquid water squeezed from the inner volume of the droplet.[16] They only occur for droplets with diameters greater than 50 µm and at temperatures higher than -25 °C.[1] This is because smaller droplets have higher surface area to volume ratios, and thus the liquid would freeze before beading. Colder environmental temperatures have the same effect. Beads were always observed to form in alignment with the vertical axis, as liquid flows in the direction of gravity.[1] However, once frozen, a droplet could reorient itself, and the bead could align with any arbitrary axis.

Turning to the internal freezing of the bulk, a large bubble often formed at the center of the droplet. Freezing generally proceeded from the outside of the droplet towards the center. As the solid front grew, the concentration of dissolved air was reduced in the ice phase and enriched in the liquid phase, bubbles then form as air is rejected from the crystal lattice.[3, 14] This is because the solubility of air in ice is three orders of magnitude smaller than in liquid water.[46] Due to the growth direction of the solid phase, bubbles were always found at the droplet center. Moreover, the size of these bubbles is usually related to heat transport and the velocity at which the droplets freeze.[3, 47] However, the freezing rates at every position were similar, and there was no evidence to suggest that bubble formation was more significant at any position. In other words, if the diffusive transport within the droplets was modified by the convective contribution of the acoustic field, the changes were similar at each position. The presence of multiple droplets was also not observed to affect bubble formation. The bubbles could vary between a single large bubble or multiple smaller bubbles. This number depends on the type of ice network created inside the droplet and the temperature exchange at the droplet surface, which is akin to the freezing rate.[1] Images of bubble formation are available in the supporting material.

3.2.2 Completely Frozen Droplets: Morphology, Sphericity, and Volumetric Expansion

Bulk freezing ended when no liquid water remained within the droplet. A majority of the completely frozen droplets consisted of a frozen core, made entirely of ice but often with a central, ellipsoidal air pocket, surrounded by surface protrusions. Notably, the protrusions create an insulating layer that allows a more significant portion of the droplet to reach colder temperatures. Additionally, the final protrusions are of low density and, therefore, likely do not affect the droplet's stability or surrounding droplets if they are long enough, though they may determine the axis of rotation in the acoustic field. Large protrusions such as these were always perpendicular to the axis of rotation. In two-droplet systems, an additional ice body formed from ice particles in an initially empty node could determine the final protrusion length. As this body grew, it may have released secondary ice particles. This increased the final amount and length of protrusions on the droplets in the surrounding positions. Conversely, if a third body did not form, very few or smaller protrusions formed on the surrounding droplets. It was discussed in the previous section that there was greater protrusion formation in three-droplet systems. INPs in two-droplet systems might have acted as a third droplet, specifically in the role of an additional source of ice. Therefore, it is also possible that atmospheric particles could add protrusions to atmospheric water droplets, and multiple droplets in proximity are not necessary for significant protrusion growth.

The volumetric expansion of water during phase change to ice is estimated to have a value of approximately 9%.[48] In this study, the same expansion was measured to be 30.75% on average. Note that protrusions were assumed not to be part of the frozen volume; only the ice core was taken. This did, however, include the frozen bead if it formed. The lowest measured expansion was for the single position 2 droplet (19.29%), though the other single droplets or those in the other configurations hued closer to the average. For instance, the highest expansion measured (in position 3 of the 1/3 configuration) was 35.61%. The higher volumetric expansion in levitated droplets may be influenced mainly by solution purity and droplet oscillation. The 9% volume expansion approximation is based on the crystal lattice structure of pure, sessile water. However, the water in this study contains entrained air trapped in the crystal structure, mainly in the center. This air is included as part of the frozen core volume. Moreover, droplet oscillations in the liquid state may cause significant defects in the crystal structure during solidification by frequently modifying the kinetics at the phase boundary. These effects were not measurable in this study and require further investigation. Therefore, the effective density of the ice core may be less than the expected density of pure ice in the same atmospheric conditions and result in a significantly greater expansion.[6, 10] It is notable that position 2, the most stable position, had the lowest expansion. This could further indicate that droplet oscillation contributes to the crystal volume increases. No other significant positional or multi-drop effects on volumetric expansion were observed.

The sphericity of a frozen droplet can be defined as the surface area of a volume-equivalent sphere divided by the measured surface area of the particle.[49] As in volumetric expansion measurements, protrusions were not included, but frozen beads were; bead formation was a significant factor in reducing droplet sphericity. For individual droplets, sphericity followed the distance from the cryogun. The highest sphericity was at position 1 (0.941), where the cooling rate was highest, and the droplet had the least time to form a significant bead. For two-droplet systems, the position 2 droplet had significant reductions in sphericity: from 0.822 individually to 0.733. No reduction was observed in the 1/3 position where the droplets are furthest apart. This follows the nucleation time trend and may indicate that sphericity is affected by the presence of INPs, and thus protrusion growth. Furthermore, there was a significant reduction in sphericity in the triple-droplet system at all positions, the average sphericity in this system was 0.645; as crystallization progresses and protrusions become increasingly prominent, the droplets continuously lose sphericity. As discussed, the three-droplet system was observed to have the greatest amount of protrusion formation, likely due to the higher available concentration of ice particles in the gas surrounding the droplets. This may demonstrate again that INP concentration, and thus protrusion formation, affects the shape of the droplet surface and, in turn, reduces droplet sphericity.

3.3 Droplet Melting

After three minutes of cooling, the cryogun was removed for the single position 2 droplets, and the solid droplets were allowed to melt from exposure to room temperature, such as in **Figure 5**. Almost instantaneously, the protrusions would melt, and the droplet would lose part of its structural, surface eccentricity and become more oblate. Soon after, convection currents would develop, and the droplet continued to melt as it spun about the lateral axis. The movement would cease once no solid ice was left, and the droplet would return to its initial temperature. The effects demonstrated in **Figure 5** are expanded upon and explained in the following sections.

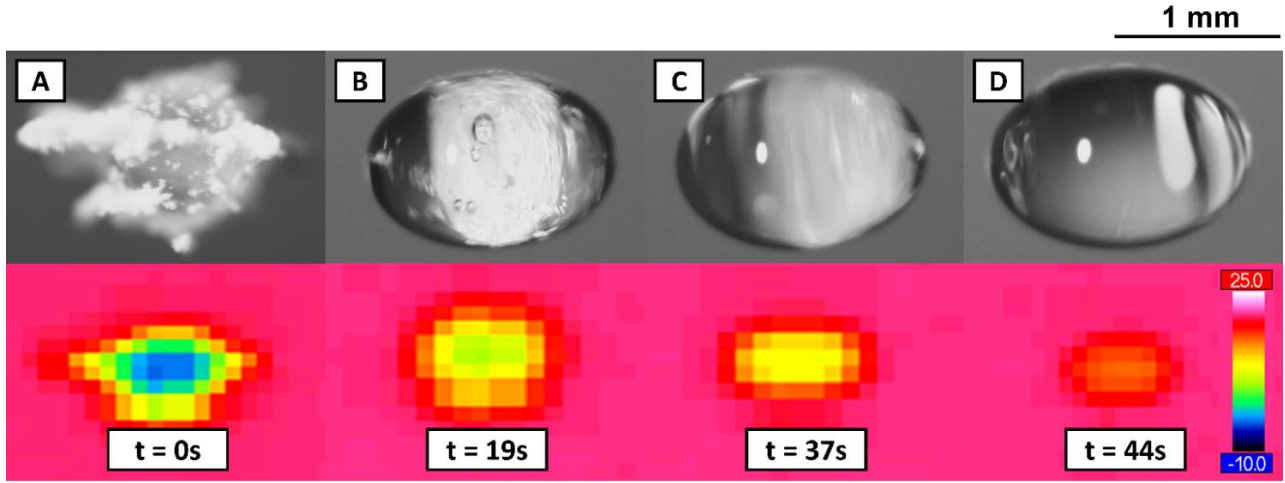

**Figure 5.** The stages of droplet melting include a completely frozen droplet (A), loss of outer ice structure (B), formation of intra-droplet currents (C), and a completely liquid final droplet (D).

3.3.1 Loss of Structure and Intra-Droplet Currents

After the freezing time had elapsed, the cooling stream was removed, and the wholly frozen position 2 droplets were exposed to the room air at 21.5 °C. Within one second, the protrusions liquified, surrounding the ice core. Likely, they liquified quickly due to their high surface area and low density; these factors resulted in a rapid temperature increase. In the same time frame, the droplet began to return to the oblate spheroid shape, though the significant amount of ice left in the droplet left a more angular appearance, such as in **Figure 5**B. It is unlikely that the protrusions alone contained significant liquid volume for the immediate, observable return of this shape around the remaining ice core. Therefore, as soon as the cooling stream is removed, some of that core's surface may also melt due to the sudden, large temperature gradient created at the gas-liquid interface. After approximately 25 seconds, the remaining ice would spin rapidly about the horizontal axis until it had melted entirely, such as in **Figure 5**C. The flow effects in a melting droplet for both a levitated and non-levitated droplet are presented in **Figure 6**. In both cases, the buoyant force brings the ice to the top of the droplet and creates a thermal gradient: the top of the droplet remains closer to the ice temperature, while the solid or gaseous interface warms the bottom. This temperature gradient results in the formation of convective currents which originate from the droplet center, its coldest portion, and move towards the bottom. The currents then form convective cells by creating a shear flow along the sides of the droplet, due to surface tension gradients along the air-liquid interface, until returning to the center.[50] Convection due to thermal gradients is called Bénard-Marangoni convection, and its presence can be suggested by calculating the Marangoni number (Ma) as in equation (5):

$$Ma = -\left(\frac{\partial \gamma}{\partial T}\right)\frac{L\Delta T}{\mu \alpha} \tag{5}$$

Where $\frac{\partial \gamma}{\partial T}$ is the surface tension rate of change with temperature (-0.154 mN/(m K), L is the liquid layer thickness (0.568 x $10^{-3}$ m on average), $\Delta T$ is the temperature difference from the droplet center to the edge (21.50 K), $\mu$ is the average viscosity (1.433 mPa s), and $\alpha$ is the average thermal diffusivity (1.43 x $10^{-7}$ m$^2$/s).[51] When Ma is large, convection flow occurs driven by the surface tension gradients. In this study, Ma is 1.89 x $10^4$, which is considered large and indicates that the convection cells produced during melting are driven by the thermal gradient. However, when a surface is present, the ice body does not rotate around the droplet. In levitated droplets, it may be that the lack of a solid interface reduces the resistance to flow along the convective path, and the currents reach the ice with a higher velocity. This may be sufficient for the ice to be caught in the convective cell, causing it to spin until it is completely melted, and the thermal gradient is no longer sufficiently large. It may also be that there is a greater thermal gradient in levitated droplets than in sessile droplets, which would enhance the thermocapillary effect. A levitated droplet is subject to a large thermal gradient over its entire surface, whereas sessile droplets have lower thermal gradients at the solid-liquid interface. Therefore, the thermal gradients, and thus the surface tensions gradients, are larger and the convective cells are stronger. Indeed, the Marangoni number for sessile systems with the same drop size and similar temperature gradients has previously been calculated as $1.58 \times 10^4$ which is much smaller than the levitated Ma, though still sufficient to induce convection.[50] However, the Marangoni numbers of both systems are fairly similar: if melting occurred in the sessile system just from turning off the cold plate, then the thermal gradients would be different, but these plates can be heated to produce the same thermal gradients. Therefore, it is more likely that the solid surface is applying some resistance to the flow.

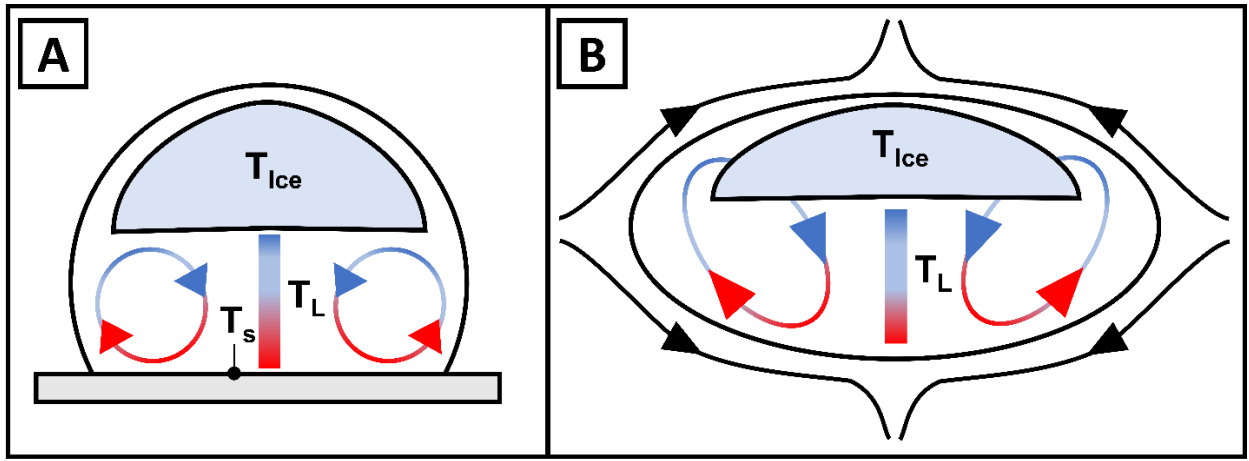

**Figure 6.** Thermal gradients cause convection currents in melting droplets. The currents stay below the remaining ice in a sessile droplet on a surface (A) but have greater mobility and can move the remaining ice in a levitated droplet (B) which induces rotation.

### 3.3.2 Final Droplet Conditions

Melting concluded when the droplet reached about 17 °C. The average final measured volume was 1.25 ± 0.15 µL (the average initial volume of 1.38 ± 0.23 µL). This could mean that there was a slight volume reduction. This reduction could come from mass losses during freezing, for instance, protrusions dislodged from the droplet by the acoustic field or from evaporation due to acoustic streaming before freezing and during melting. It was expected that, as nucleation and, critically, protrusion formation may be caused by secondary ice particles, there would be a slight increase in the final volume. This could indicate that protrusion growth is primarily fed by the initial volume and not the further agglomeration of INPs. Equally, volume additions from this process may be less significant than the losses from evaporation processes. However, the overall impact of these processes is difficult to determine. The average change in volume (including a confidence interval) is -0.13 ± 0.14 µL: there was no statistically significant volume change. Therefore, this study could not determine absolute volume gain and loss values. Evaporation losses for droplets of similar size have previously been investigated and found to be approximately 1.5% of the initial volume.[18, 27] Assuming evaporation occurs over the entire experimental time, this is equivalent to 0.021 µL. This could play a role in the negative average change in volume, though it is likely not significant. Therefore, it is more likely that any mass losses came from dislodging the protrusions that were partially made from the initial liquid bulk.

4. CONCLUSIONS

Previous examinations of phase change in levitated water have been limited to single-droplet, single-position studies, often in a stationary acoustic field contained by a cooling chamber.[1, 3, 5, 11-13, 44] Building on this strong base of work, pure water droplets were frozen and melted while acoustically levitated in three positions to determine the presence of inter-droplet effects during phase change. Concurrent digital and thermal images were directly taken of single droplets and double- and triple-droplet configurations. This was the first time multiple levitated water droplets were frozen and examined in this manner as previous levitation techniques did not allow for several droplets to crystalize at once and cooling chambers only allowed for limited visualization and no IR imaging. Additionally, levitated melting had yet to be sufficiently examined. The liquid droplets initially adopted an oblate spheroid shape in the acoustic field which was the net result of streaming forces and surface tension at the acoustic boundary layer (the air-liquid interface). When the cooling stream was introduced, however, the additional, counteracting vertical force at the interface resulted in the circularization of the droplet. Further examination of hydrostatic forces and internal viscous flow at this interface on droplet geometry are suggested for future work. Nucleation in the droplets was determined to be a function of cooling rate and not undercooling, which indicated that it was secondary rather than primary. Secondary ice-nucleating particles came either from aerosolized atmospheric water or were produced from droplets that had already nucleated, the latter mechanism for which was mainly droplet breakup. This was where there was sufficient energy from increased internal droplet pressure to overcome capillary forces and negative pressure inside the droplet, resulting in rupture at the solid-liquid interface. In the presence of other droplets, the droplet at position 2 exhibited a significant reduction in nucleation time due to the increased concentration of INPs in the surrounding gas. There was no statistical difference in the nucleation times of droplets in adjacent positions. Therefore, the initial solid-liquid interfacial fracture of one droplet resulted in the nucleation of another. During bulk crystal growth, one or more protrusions of varying thickness could form on the surface of the droplet; also initiated by the same INPs that caused nucleation. This was possible as a quasi-liquid layer formed between the solid and gaseous interfaces. The location and shape of the protrusions were determined by the local interfacial acoustic field forces and changes in surface tension. When the ice core was completely frozen, an ellipsoidal air bubble was often found at its center, pushed there by the solid front and created due to the lower solubility of air in ice compared to water at the internal solid-liquid interface. Due to the presence of this air and oscillations at the phase boundary, which also require further study to be accurately characterized, volumetric expansion was found to be 30.75% compared to 9% in pure, sessile water. Melting began with the rapid loss of solid structure and a return to the oblate spheroidal droplet shape. Due to the buoyancy of ice and thermocapillary effects along the gas-liquid interface, intra-droplet convective currents were created, and the droplet spun while melting. This study shows that there exist inter-droplet effects between nucleating and crystallizing water particles which affect both the droplet phase transition and morphology. These new insights go beyond the freezing of individual droplets and may further our current understanding of collective atmospheric crystallization phenomena. Future studies should consider levitated colloidal suspensions with lower surface tensions that could change the interfacial crystallization characteristics or levitated hydrates which would modify the structure of the initial solid-liquid interface.

AUTHOR INFORMATION

**Corresponding Author**

*phillip.servio@mcgill.ca

**Author Contributions**

The manuscript was written through the contributions of all authors. All authors have given approval to the final version of the manuscript.

ACKNOWLEDGEMENTS

The authors would like to acknowledge the financial support from the Natural Sciences and Engineering Research Council of Canada (NSERC) and the Faculty of Engineering of McGill University (MEDA, Vadasz Scholars Program).

SUPPORTING MATERIAL

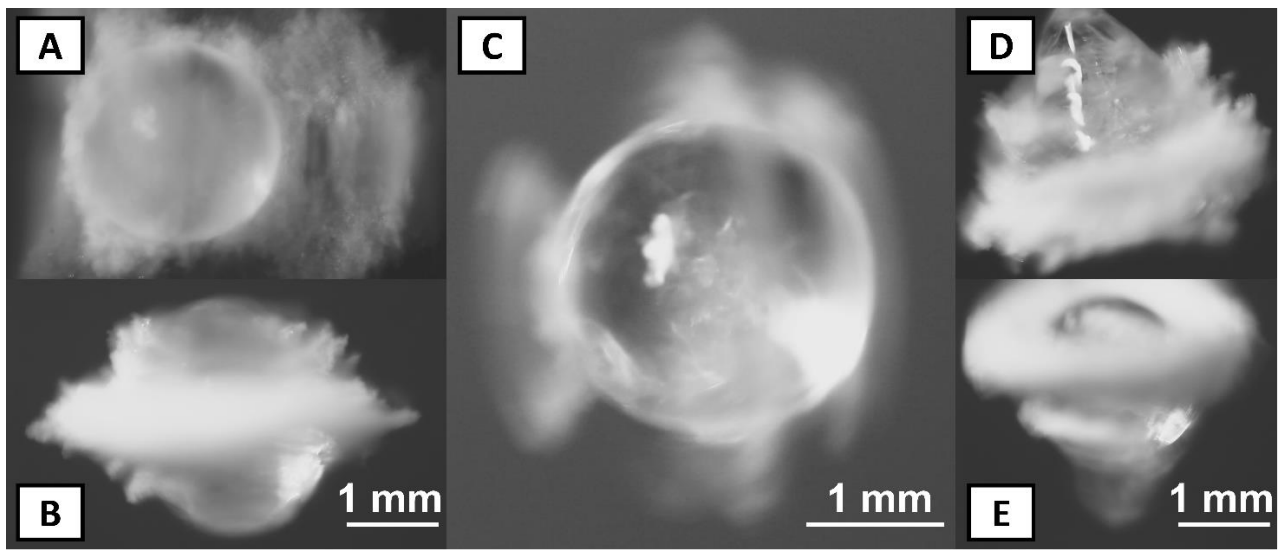

**Figure S1.** Different types of protrusion such as thin and wisp-like (A), a single, thick band (B), or several thinner bands (C). Protrusion formation did not affect stability; a droplet could remain stable regardless of orientation (D and E).

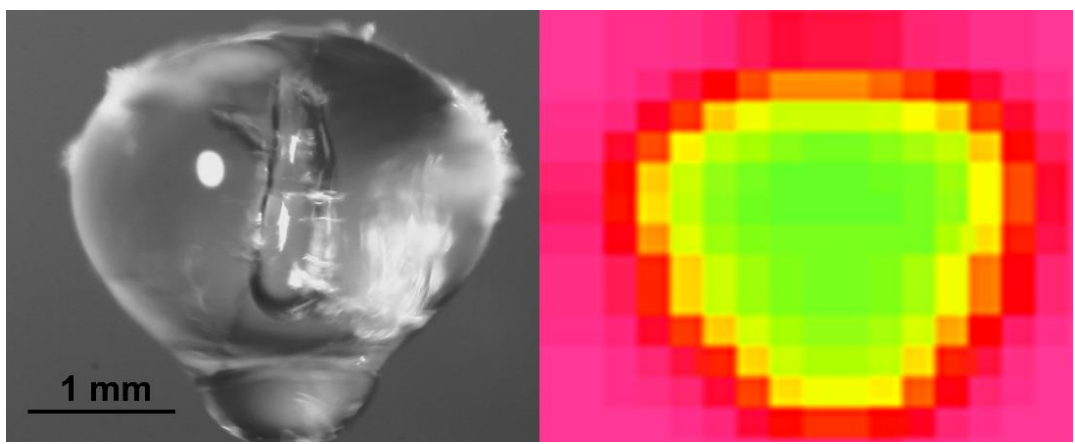

**Figure S2.** Digital (left) and thermal (right) images of a droplet with a frozen bead. This bead was significant enough to change the droplet shape in the IR imaging.

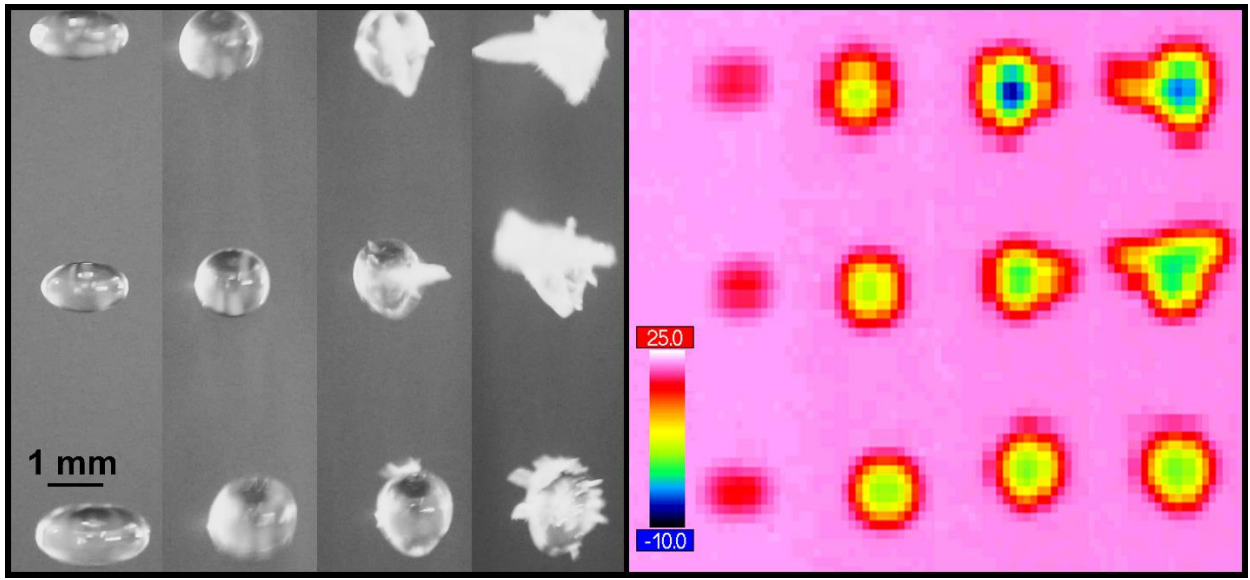

**Figure S3.** Digital (left) and thermal (right) images of freezing in the triple-droplet system from the initial droplets to nucleation, bulk growth, and complete solidification.

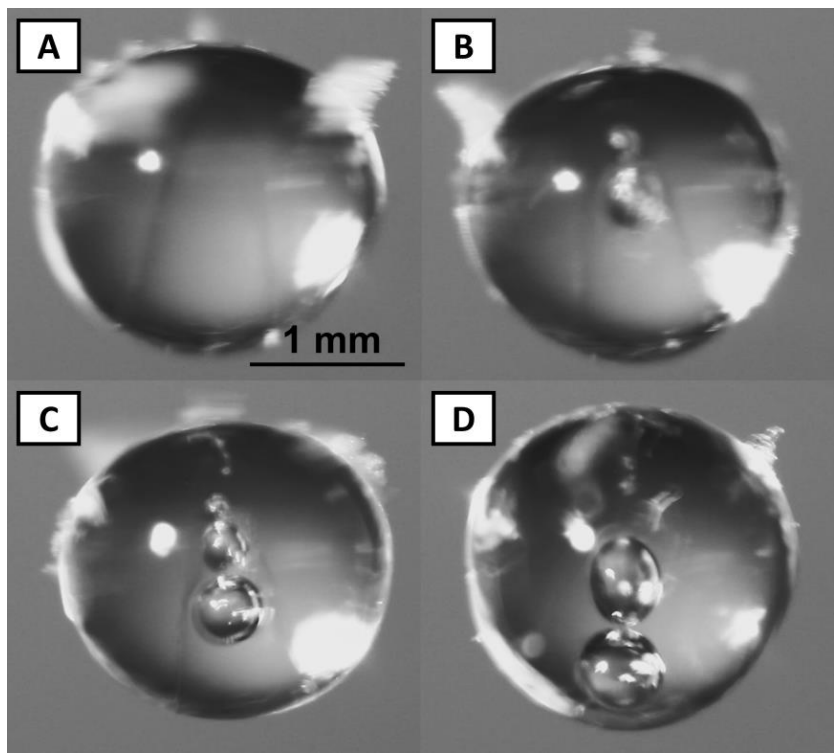

**Figure S4.** Bubble formation during bulk freezing. Initially, no bubbles are present (A), then bubbles form in the liquid in the smallest depression of the solid front (B), bubbles continue to form as the drop further solidifies (C), and eventually, there is no liquid left, and the bubbles become air pockets in a solid drop (D).